\documentclass[11pt,a4paper,english,superscriptaddress,prd]{revtex4}
\usepackage{amssymb}
\usepackage{amsmath} 
\usepackage{slashed}
\usepackage{graphicx}
\usepackage{babel}
\usepackage{hyperref}
\usepackage{color}
\usepackage{comment}
\allowdisplaybreaks[1]
\makeatletter\AtBeginDocument{\let\@elt\relax}\makeatother

\begin{document}

\title{Gravitational corrections to the Einstein-Scalar-QCD model}

\author{Huan Souza}
\email{huan.souza@icen.ufpa.br}
\affiliation{Faculdade de F\'isica, Universidade Federal do Par\'a, 66075-110, Bel\'em, Par\'a, Brazil.}

\author{L.~Ibiapina~Bevilaqua} \email{leandro.bevilaqua@ufrn.br} \affiliation{Escola de Ci\^encias e Tecnologia, Universidade Federal do Rio Grande do Norte\\ Caixa Postal 1524, 59072-970, Natal, Rio Grande do Norte, Brazil.}

\author{A.~C.~Lehum}
\email{lehum@ufpa.br}
\affiliation{Faculdade de F\'isica, Universidade Federal do Par\'a, 66075-110, Bel\'em, Par\'a, Brazil.}

\begin{abstract}
This study employs the effective field theory approach to quantum gravity to investigate a non-Abelian gauge theory involving scalar particles coupled to gravity. The study demonstrates explicitly that the Slavnov-Taylor identities are maintained at one-loop order, which indicates that the universality of the color charge is preserved. Additionally, the graviton corrections to the two-loop gluon self-energy and its renormalization are computed.

\end{abstract}

\maketitle

\section{Introduction}

Although we are still in need of a consistent and generally accepted description of quantum gravity at high energies, if we restrict ourselves to low energies compared to the Planck scale, we can nevertheless draw some trustful conclusions about the gravitational phenomena at quantum level using the viewpoint and methods of effective field theories \cite{Donoghue:1994dn, Burgess:2003jk, Shapiro}. Thus, the well known nonrenormalizability of Einstein's theory coupled to other fields \cite{'tHooft:1974bx, PhysRevLett.32.245, Deser:1974cy} is not an impediment to study the influence of gravity in the renormalization of other fields and parameters in a meaningful way. The central idea is that we add to the action the high-order terms needed to renormalize the parameters of the lower-order terms and the new parameters introduced will be irrelevant to the low-energy behavior of the theory.

As it is well known, the renormalized quantities of a theory depend on an arbitrary scale and the renormalization group is the theoretical tool to study this dependence and allows us to describe how the coupling constants change with this scale, establishing the so-called running of the coupling constants \cite{Srednicki:2007qs}. If this dependence is such that the coupling constant gets weaker as we go to higher energies the theory is said to be asymptotically free \cite{Gross:1973id, Politzer:1973fx, Gross:1974jv}. The possibility that gravitational corrections could render all gauge coupling constants asymptotically free was suggested by Robinson and Wilczek, who used the effective field theory approach of quantum gravity to reach this conclusion \cite{Robinson:2005fj}. However, this result was soon contested by Pietrykowski \cite{Pietrykowski:2006xy}, who showed that the result was gauge dependent. Subsequently, many works investigate the use of the renormalization group in quantum gravity as an effective field theory (See for instance Refs.~\cite{Felipe:2012vq,Felipe:2013vq,Ebert:2007gf,Nielsen:2012fm,Toms:2008dq,Toms:2010vy,Ellis:2010rw,Anber:2010uj,Bevilaqua:2015hma,Bevilaqua:2021uzk,Bevilaqua:2021uev}). In a previous work \cite{Bevilaqua:2015hma}, we used dimensional regularization to compute gravitational effects on the beta function of the scalar quantum electrodynamics at one-loop order and found that all gravitational contributions cancel out. The situation is different at two-loop order, in which we do find nonzero gravitational corrections to the beta function for both scalar and fermionic QED, as shown in a latter work \cite{Bevilaqua:2021uzk}. However, those corrections give a positive contribution to the beta function and thus the electrical charge is not asymptotically free neither has a nontrivial fixed point.

The use of renormalization group in the context of non-renormalizable field theories raise some subtle questions. The universality of the coupling constants in effective field theories was discussed by Anber \textit{et al.} in \cite{Anber:2010uj}, where it was suggested that an operator mixing could make the coupling constants dependent on the process under consideration and therefore non-universal. That would imply that, unlike renormalizable field theories, the concept of running coupling may not be useful in the effective field theory approach to quantum gravity. This is indeed the case for the quartic self-interaction of scalars in scalar-QED, as discussed in \cite{Bevilaqua:2015hma} but, as shown in \cite{Bevilaqua:2015hma} for scalar-QED and in \cite{Bevilaqua:2021uev} for fermionic-QED it seems not to be the case for the gauge coupling because of the Ward identity. The central role of the gauge symmetry in the universality of the gauge coupling for QED led us to explore this issue in the non-Abelian case. Using dimensional regularization, we showed that the Slavnov-Taylor identities are satisfied in a non-Abelian gauge theory coupled to fermions and gravity \cite{Souza:2022ovu}. In the same work, we have also calculated the gravitatinal correction for the beta function at one-loop thus verifying directly the absence of contributions from the gravitational sector.

In previous studies, the coupling of non-Abelian gauge theories to gravity has been investigated \cite{Souza:2022ovu,Buchbinder:1983nug,Tang:2008ah,Tang:2011gz}. In this research, we extend our previous analysis by investigating the asymptotic behavior of a non-Abelian gauge theory coupled to complex scalars and gravity. This exploration is motivated by the significant role scalar theories play in the advancement of high-energy theory. Over the years, scalar models have been proposed to tackle issues such as renormalization group theory for non-renormalizable theories \cite{Barvinsky:1993zg}, the study of dilatons \cite{Shapiro:1995yc}, and potential candidates for dark matter \cite{Cohen:2011ec, Arkani-Hamed:2008hhe}.. In fact, Ref.~\cite{Calmet:2021iid} argue that quantum gravity might have crucial implications in a theory of dark matter. Additionally, a recent study \cite{and:2022ttn} investigated the interaction between SU(2) Yang-Mills waves and gravitational waves. The results revealed that while the problem can be perturbatively studied in the symmetric phase, non-perturbative approaches are necessary in the broken phase. Hence, the examination of a non-Abelian gauge theory coupled to complex scalars and gravity is of particular interest due to the fundamental role scalar theories have played in addressing diverse problems in high-energy theory.

The paper is structured as follows. Section \ref{sec2} introduces the Lagrangian and propagators of the model. In Section \ref{sec3}, the one-loop renormalization of the model is presented, highlighting the preservation of gauge invariance of the gravitational interaction and respect for the Slavnov-Taylor identities. Section \ref{sec4} utilizes the Tarasov algorithm to compute the two-loop counterterm for the gluon wave-function. Finally, concluding remarks are provided in Section \ref{summary}. The minimal subtraction (MS) scheme is used throughout this work to handle the UV divergences, with $(+---)$ being the spacetime signature and natural units of $c=\hbar=1$ are adopted.

\section{The Einstein-Scalar-QCD model}\label{sec2}

To get an effective field theory description for our model, we add higher order terms to the Lagrangian of a non-Abelian gauge theory with complex scalars coupled to gravity:
\begin{eqnarray}\label{fQCD}
\mathcal{L}=&& \sqrt{-g}\sum_f\Big\{\frac{2}{\kappa^2}R-\frac{1}{4} g^{\mu\alpha}g^{\nu\beta} G_{\mu\nu}^a G_{\alpha\beta}^a + g^{\mu\nu}(D_{\mu}\phi^i)^\dagger D_\nu \phi^i - m_i(\phi^i)^\dagger\phi^i +\lambda((\phi^i)^\dagger\phi^i)^2+\mathcal{L}_{HO}\Big\},\nonumber\\
&&~
\end{eqnarray}
\noindent where the index $i=1,2,\cdots,N_s$ runs over the scalars flavors, $G^a_{\mu\nu}=\nabla_\mu A_\nu^a-\nabla_\nu A_\mu^a + gf^{abc}A^b_\mu A^c_\nu$ is the non-Abelian field-strength with $f^{abc}$ being the structure constants of the $SU(N)$ group, and $D_\mu = \partial_\mu -ig t^aA^a_\mu$ is the covariant derivative. The higher order terms $\mathcal{L}_{HO}$ are written as
\begin{equation}
 \mathcal{L}_{HO} = \frac{\tilde{\lambda}_1}{M_P^2}\left[\mathrm{Re}((\phi^i)^\dagger\partial_\mu\phi^i)\right]^2
  +\frac{\tilde{\lambda}_2}{M_P^2}\left[\mathrm{Im}((\phi^i)^\dagger\partial_\mu\phi^i)\right]^2-\frac{\tilde{e}_3}{4}G_a^{\mu\nu}\frac{\Box}{M_P^2} G^a_{\mu\nu}.
\end{equation}

To obtain the usual quadratic term for the gravitational field, we need to expand $g_{\mu\nu}$ around the flat metric as
\begin{equation}\label{metric}
 g_{\mu\nu} = \eta_{\mu\nu} + \kappa h_{\mu\nu},
\end{equation}
such that
\begin{equation}\label{sqrt}
g^{\mu\nu} = \eta^{\mu\nu} - \kappa h^{\mu\nu} +\cdots \qquad \text{and} \qquad  \sqrt{-g} = 1 + \frac{\kappa}{2}h + \cdots,
\end{equation}
\noindent where $h = \eta^{\mu\nu}h_{\mu\nu}$. The affine connection is written as
\begin{equation}\label{connection}
 \Gamma^\lambda_{~\mu\nu} = \frac{1}{2}\kappa(\eta^{\lambda\sigma} - \kappa h^{\lambda\sigma})(\partial_\mu h_{\sigma\nu} + \partial_\nu h_{\sigma\mu} - \partial_\sigma h_{\mu\nu}).
\end{equation}

Organizing the Lagrangian as,
\begin{subequations}
 \begin{eqnarray}
  \mathcal{L} &=& \mathcal{L}_h + \mathcal{L}_f + \mathcal{L}_A;\\
  \mathcal{L}_h &=& \frac{2}{\kappa^2}\sqrt{-g}R;\label{gravsector}\\
  \mathcal{L}_s &=& \sqrt{-g}[g^{\mu\nu}(D_{\mu}\phi^i)^\dagger D_\nu \phi^i - m_i(\phi^i)^\dagger\phi^i +\lambda((\phi^i)^\dagger\phi^i)^2];\label{mattersector}\\
  \mathcal{L}_A &=& -\frac{\sqrt{-g}}{4}g^{\mu\alpha}g^{\nu\beta}G_{\mu\nu}^aG^a_{\alpha\beta}\label{gaugesector}.
 \end{eqnarray}
\end{subequations}

Using Eqs.~\eqref{metric}-\eqref{connection}, we write the pure gravity sector \eqref{gravsector} in terms of $h_{\mu\nu}$. Moreover, it is convinient to organize $\mathcal{L}_h$ in powers of $h$ as follows:
\begin{subequations}\label{Lh}
 \begin{eqnarray}
  \mathcal{L}_h &=& \mathcal{L}_h^0 + \kappa\mathcal{L}_h^1 + \cdots\\
  \mathcal{L}_h^0 &=& -\frac{1}{4}\partial_\mu h\partial^\mu h + \frac{1}{2}\partial_\mu h^{\sigma\nu}\partial^{\mu}h_{\sigma\nu};\\
  \mathcal{L}_h^1 &=& \frac{1}{2}h^\alpha_{~\beta}\partial^\mu h^\beta_{~\alpha}\partial_\mu h - \frac{1}{2}h^\alpha_{~\beta}\partial_\alpha h^\mu_{~\nu}\partial^\beta h^\nu_{~\mu} - h^\alpha_{~\beta}\partial_\mu h^\nu_{~\alpha}\partial^\mu h^\beta_{~\nu}\nonumber\\
  && +\frac{1}{4}h\partial^\beta h^\mu_{~\nu}\partial_\beta h^\nu_{~\mu} + h^\beta_{~\mu}\partial_\nu h^\alpha_{~\beta}\partial^\mu h^\nu_{~\alpha} - \frac{1}{8}h\partial^\nu h\partial_\nu h,
 \end{eqnarray}
\end{subequations}
where the indices are raised and lowered with the flat metric (here and henceforth, we are following the results in Ref. \cite{Choi:1994ax}).

For the matter sector \eqref{mattersector}, the expansion around the flat metric give us
\begin{subequations}
 \begin{eqnarray}
  \mathcal{L}_s &=& (D^\mu\phi^i)^\dagger D_\mu\phi^i - m^2_i((\phi^i)^\dagger\phi^i) -\frac{\lambda}{4}((\phi^i)^\dagger\phi^i)^2 - \kappa h^{\mu\nu}(D_\mu\phi^i)^\dagger D_\nu\phi^i\nonumber\\
  &&+ \frac{\kappa}{2}h\left[(D^\mu\phi^i)^\dagger D_\mu\phi^i - m^2_i (\phi^i)^\dagger\phi^i -\frac{\lambda}{4}((\phi^i)^\dagger\phi^i)^2\right],
 \end{eqnarray}
\end{subequations}
which we organize as follows
\begin{subequations}
 \begin{eqnarray}
  \mathcal{L}_s &=& \mathcal{L}_s^0 + \kappa \mathcal{L}_s^1 + \cdots\\
  \mathcal{L}_s^0 &=& (D^\mu\phi^i)^\dagger D_\mu\phi^i - m_i^2((\phi^i)^\dagger\phi^i) -\frac{\lambda}{4}((\phi^i)^\dagger\phi^i)^2 \\
  \mathcal{L}_s^1 &=& -h^{\mu\nu}(D_\mu\phi^i)^\dagger D_\nu\phi^i+ \frac{1}{2}h\left[(D^\mu\phi^i)^\dagger D_\mu\phi^i - m_i^2 (\phi^i)^\dagger\phi^i -\frac{\lambda}{4}((\phi^i)^\dagger\phi^i)^2\right];
 \end{eqnarray}
\end{subequations}
and finally, for the gauge sector,
\begin{subequations}\label{LA}
 \begin{eqnarray}
  \mathcal{L}_A &=& \mathcal{L}_A^0 + \kappa\mathcal{L}_A^1 + \cdots\\
  \mathcal{L}_A^0 &=& -\frac{1}{4}G_{\mu\nu}^aG^{\mu\nu}_a \\
  \mathcal{L}_A^1 &=& \frac{1}{2}h^\tau_{~\nu}G^{\mu\nu}_aG_{\mu\tau}^a + \frac{1}{2}h\mathcal{L}_A^0.
 \end{eqnarray}
\end{subequations}

As usual for gauge theories, in order to quantize this model, we have to deal with the excess of degrees of freedom in $A_\mu^a$ and $h_{\mu\nu}$ due to their symmetries. In our calculations, we have followed the Faddeev-Popov procedure that introduces gauge-fixing terms in the action that will modify the propagators of both $A_{\mu}^a$ and $h_{\mu\nu}$. Moreover, we must also introduce ghost fields for both vector and tensor fields. However, the ghost field associated with the graviton will not appear in this text because, since we are working with the one-graviton exchange approximation, the new term containing the ghosts added to the action will not contribute to the renormalization of the gauge coupling constant. Therefore, whenever we refer to ghost field in what follows, we mean the one associated with $A_{\mu}^a$. The propagators for scalars, ghosts, gluons and gravitons are given, respectively, by
\begin{subequations}\label{propagators}
\begin{eqnarray}
\Delta_s(p) &=& \frac{i}{p^2-m_a^2};\\
\Delta_{ab}(p) &=& \frac{i}{p^2}\delta_{ab};\\
\Delta^{\mu\nu}_{ab}(p) &=& \frac{i}{p^2}\left(\eta^{\mu\nu}-(1-\xi_A)\frac{p^\mu p^\nu}{p^2} \right)\delta_{ab};\\
\Delta^{\alpha\beta\mu\nu}(p) &=& \frac{i}{p^2}\left(P^{\alpha\beta\mu\nu}-(1-\xi_h)\frac{Q^{\alpha\beta\mu\nu}}{p^2}\right). 
\end{eqnarray}
\end{subequations}
The gauge-fixing parameters $\xi_A$ and $\xi_h$ will be carried out through the whole calculation, since we do not want to choose any specific gauge. The projectors $P^{\alpha\beta\mu\nu}$ and $Q^{\alpha\beta\mu\nu}$ in the graviton propagator are given by
\begin{eqnarray}
P^{\alpha\beta\mu\nu} &=&\frac{1}{2} \left(\eta^{\alpha\mu}\eta^{\beta\nu}+\eta^{\alpha\nu}\eta^{\beta\mu}-\eta^{\alpha\beta}\eta^{\mu\nu} \right);\nonumber\\
Q^{\alpha\beta\mu\nu} &=& (\eta^{\alpha\mu}p^\beta p^\nu+\eta^{\alpha\nu}p^\beta p^\mu+\eta^{\beta\mu}p^\alpha p^\nu+\eta^{\beta\nu}p^\alpha p^\mu).
\end{eqnarray}

\section{The one-loop renormalization}\label{sec3}

The Slavnov-Taylor identities are a set of relations that must be satisfied by the n-point functions to ensure the gauge independence of the observables of the theory. In this section we want to explicitly show that the Slavnov-Taylor identities are respected at one-loop order for our model. To simplify our computations, we will consider here that all the masses are the same, so we drop the index $i$. As we will see, this will not affect our final result.

We start by computing the n-point functions. Namely, the self-energy of scalar, vector and ghost fields ($\Sigma_s, \Pi^{\mu\nu}_{ab}$ and $\Sigma_{ab}$, respectively), also the scalar-gluon, ghost-gluon and gluon-gluon three-point functions ($\Gamma^{\mu}_a$, $\Gamma^\mu_{abc}$ and $\Pi^{\mu\nu\alpha}_{abc}$, respectively), the gluon four-point function ($\Gamma^{\mu\nu\rho\sigma}_{abcd}$), and finally the scalar-gluon four-point function ($\Pi^{\mu\nu}_{abcd}$). All the computations were done using the \textit{Mathematica} packages: \textit{FeynRules} to generate the models \cite{feynrules}, \textit{FeynArts} to draw the diagrams \cite{Hahn:2000kx}, and \textit{FeynCalc} to simplify and compute the amplitudes \cite{Shtabovenko:2020gxv}.

\begin{figure}[t]
	\begin{center}
	\includegraphics[angle=0 ,width=12cm]{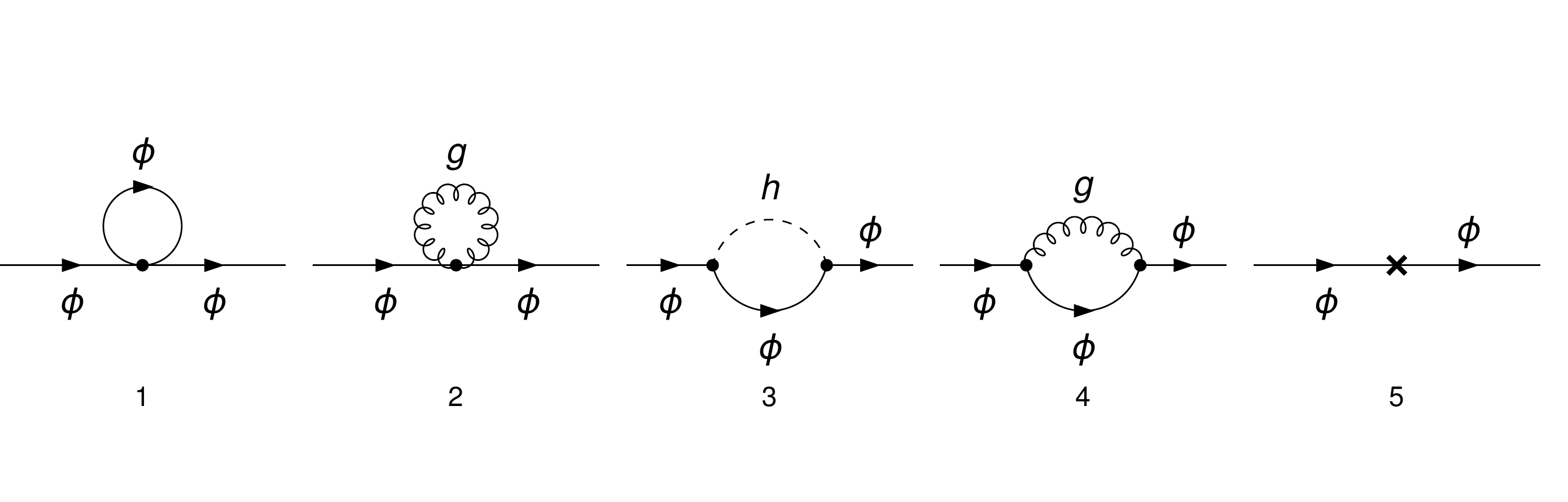}
	\caption{Feynman diagrams for the scalar self-energy. Continuous, wiggly, dotted, and dashed lines represent the scalar, gluon, ghost, and graviton propagators, respectively.}	\label{fig01}
	\end{center}
\end{figure}

At one-loop, the self-energy of the scalar field, Fig.~\ref{fig01}, results in
\begin{eqnarray}\label{1eq02}
-i\Sigma_s(p) &&= i p^2\left(\frac{C_A \left(\xi _A-3\right) g^2-(\xi_h-2)\kappa ^2 m^2}{16 \pi ^2 \epsilon } +   Z_{2s}^{(1)}\right)\nonumber\\
&&+i m^2\left(\frac{ -C_A \xi _A g^2+4 \lambda  N_s-(\xi_h-2)\kappa ^2 m^2}{16 \pi ^2 \epsilon}-Z_{m_s}^{(1)}\right) + \mathrm{finite},
\end{eqnarray}
\noindent where $C_A=N$ for the $SU(N)$ group. By imposing finiteness to $\Sigma_s(p)$, we find the following one-loop counterterms:
\begin{subequations}\label{ct01}
\begin{eqnarray}
Z_{2s}^{(1)} &=& \frac{\kappa ^2 m^2 \left(\xi _h-2\right)-C_A \left(\xi _A-3\right) g^2}{16 \pi ^2 \epsilon },\label{eq_z2f}\\
Z_{m}^{(1)} &=& \frac{ -C_A \xi _A g^2+4 \lambda  N_s-(\xi_h-2)\kappa ^2 m^2}{16 \pi ^2 \epsilon}.\label{ct01-2}
\end{eqnarray}
\end{subequations}

\begin{figure}[t!]
	\begin{center}
	\includegraphics[angle=0 ,width=10cm]{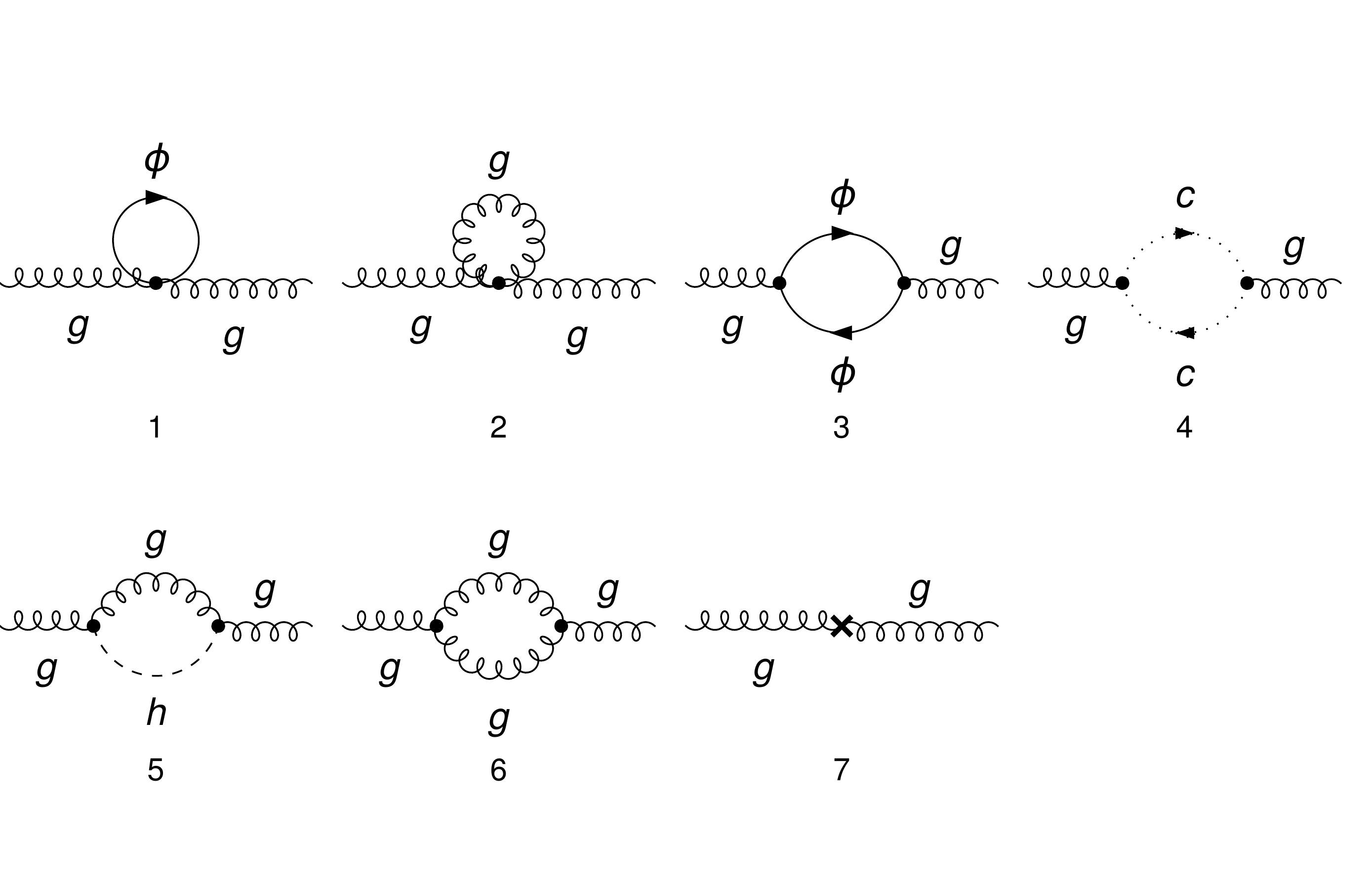}
	\caption{Feynman diagrams for the gluon self-energy.}
	\label{fig02}
	\end{center}
\end{figure}

For the gluon self-energy, it is convenient to write the one-loop correction (corresponding to the diagrams in Fig.~\ref{fig02}) as
\begin{eqnarray}\label{1eq02b}
\Pi^{\mu\nu}_{ab}(p)= \left(p^2 \eta^{\mu  \nu }-p^{\mu } p^{\nu }\right)\Pi(p)\delta_{ab},
\end{eqnarray}
\noindent where the function $\Pi(p)$ is found to be
\begin{eqnarray}\label{eq_pi1}
\Pi(p)= -iZ_3^{(1)}-ip^2\tilde{Z}_3^{(1)}+\frac{i \kappa ^2 p^2 \left(2-3 \xi _h\right)}{96 \pi ^2 \epsilon }-\frac{i  C_A g^2 \left(2 N_s+3 \xi _A-13\right)}{96 \pi ^2 \epsilon }+\mathrm{finite},
\end{eqnarray}
and, imposing the finiteness on $\Pi(p)$, we find 
\begin{subequations}\label{ct02}
\begin{eqnarray}
Z_3^{(1)} &=& -\frac{C_A g^2 \left(2 N_s+3 \xi _A-13\right)}{96 \pi ^2 \epsilon },\label{eq_z3}\\
\tilde{Z}_3^{(1)} &=& -\frac{\kappa^2(3\xi_h-2)}{96\pi^2\epsilon}~.\label{ct02-2}
\end{eqnarray}
\end{subequations}
We can see from Eq.~\eqref{eq_pi1} that $Z_3$ is the relevant counterterm to the beta function of the color charge, since it is the renormalizing factor for the quadratic term $G^{\mu\nu}_a G_{\mu\nu}^a$, while $\tilde{Z}_3$ renormalizes a higher derivative term like $G^{\mu\nu}_a\Box G_{\mu\nu}^a$. Notice also that the UV divergent part of Eq.~\eqref{eq_pi1} is not dependent on the masses of the scalars.

\begin{figure}[h!]
	\begin{center}
	\includegraphics[angle=0 ,width=5cm]{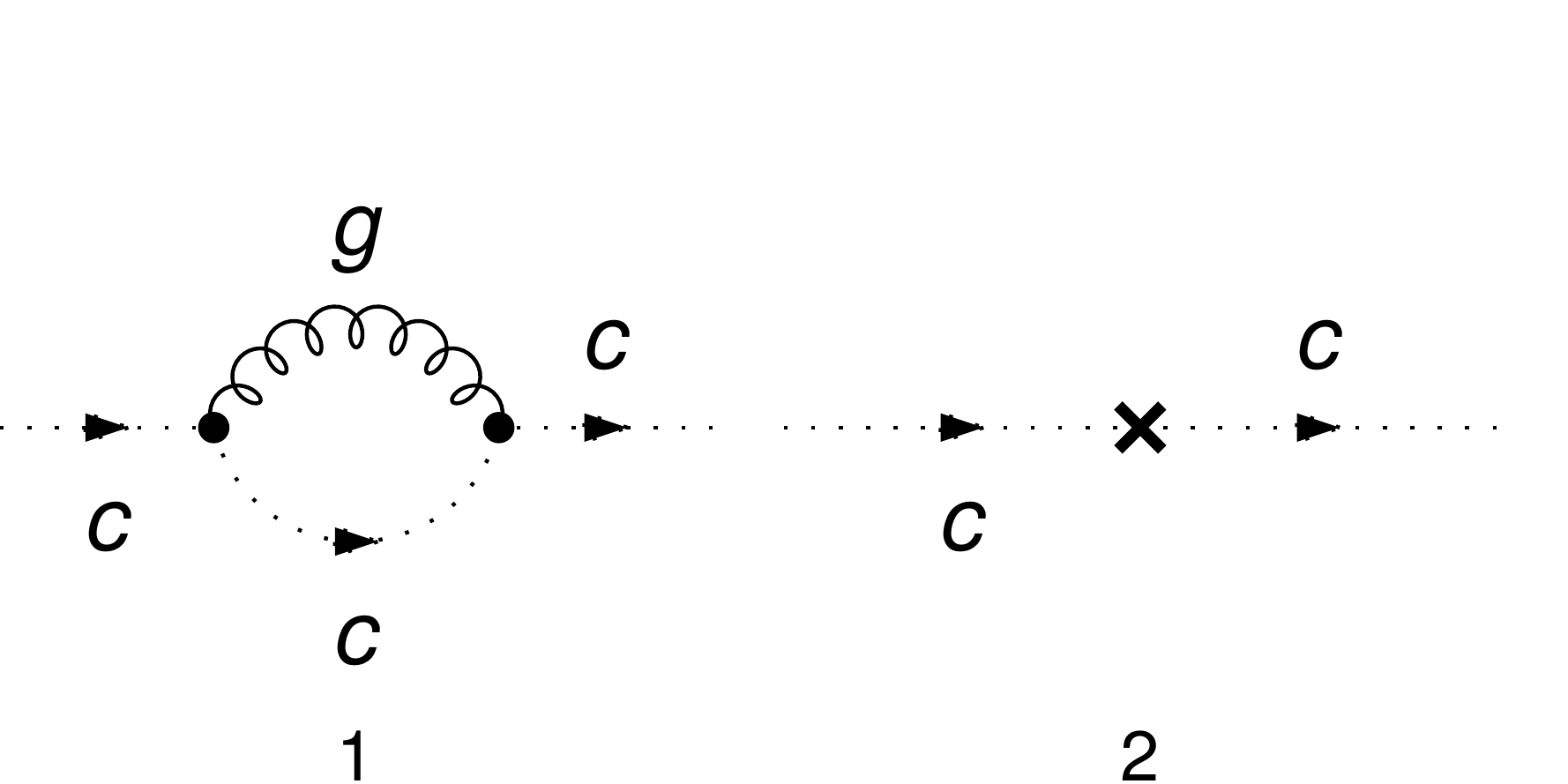}
	\caption{Feynman diagrams for the ghost self-energy.}
	\label{fig03}
	\end{center}
\end{figure}

Contributions to the ghost self-energy up to one-loop order are depicted in Fig.~\ref{fig03}. The resulting expression is
\begin{eqnarray}\label{ghostSE}
-i\Sigma_{ab} &=& \left(\frac{i p^2 C_A \left(\xi _{A}-3\right) g^2}{64 \pi ^2 \epsilon }+i p^2 Z_{2c}^{(1)}\right)\delta_{ab} +\mathrm{finite},
\end{eqnarray}
and, imposing finiteness, we find
\begin{eqnarray}\label{eq_z2c}
Z_{2_c}^{(1)} &=& -\frac{C_A g^2 \left(\xi _A-3\right)}{64 \pi ^2 \epsilon }.
\end{eqnarray}
Notice that in Fig.~\ref{fig03} the gravitational interactions are not shown. Although in the action there is a coupling of $h^{\mu\nu}$ to the kinetic term of the ghosts associated with the gluons, the gravitational contributions to the ghost self-energy will be renormalized by a higher-order term and is therefore irrelevant for our purposes here. One way to see why this is happens is to observe that both the ghosts and the graviton are massless, so the only contribution proportional to $\kappa^2$ must be of the order $p^4$.
\begin{figure}[t!]
	\begin{center}
	\includegraphics[angle=0 ,width=8cm]{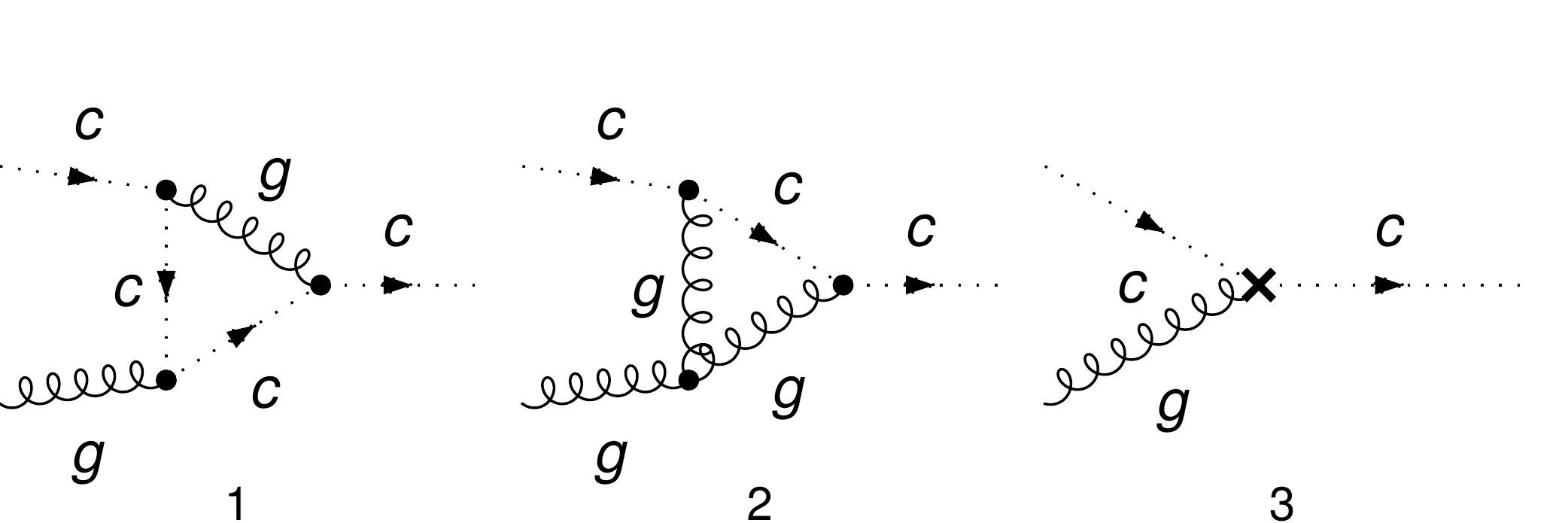}
	\caption{Feynman diagrams for the vertex interaction between gluons and ghosts up to one-loop order.}
	\label{fig04}
	\end{center}
\end{figure}

For the 3-point functions, let's first consider the ghost-ghost-gluon vertex (Fig.~\ref{fig04}), where again all the gravitational corrections are renormalized by higher-order terms and are therefore omitted here. Also, in the following expressions, we will use $p_1$ and $p_2$ to represent incoming external momenta, and $p_3$ and $p_4$ for outgoing momenta. The expression obtained for these diagrams is
\begin{equation}
 \Gamma_{abc}^\mu = -g p_3^{\mu } f_{abc}\left(\frac{C_A g^2\xi_A}{32 \pi ^2 \epsilon } + Z_{1c}^{(1)}\right) + \text{finite},
\end{equation}
and the subtraction of the UV pole will give us
\begin{equation}\label{eq_z1c}
 Z_{1_c}^{(1)} = -\frac{C_A g^2\xi_A}{32 \pi ^2 \epsilon }.
\end{equation}

\begin{figure}[t!]
	\begin{center}
	\includegraphics[angle=0 ,width=10cm]{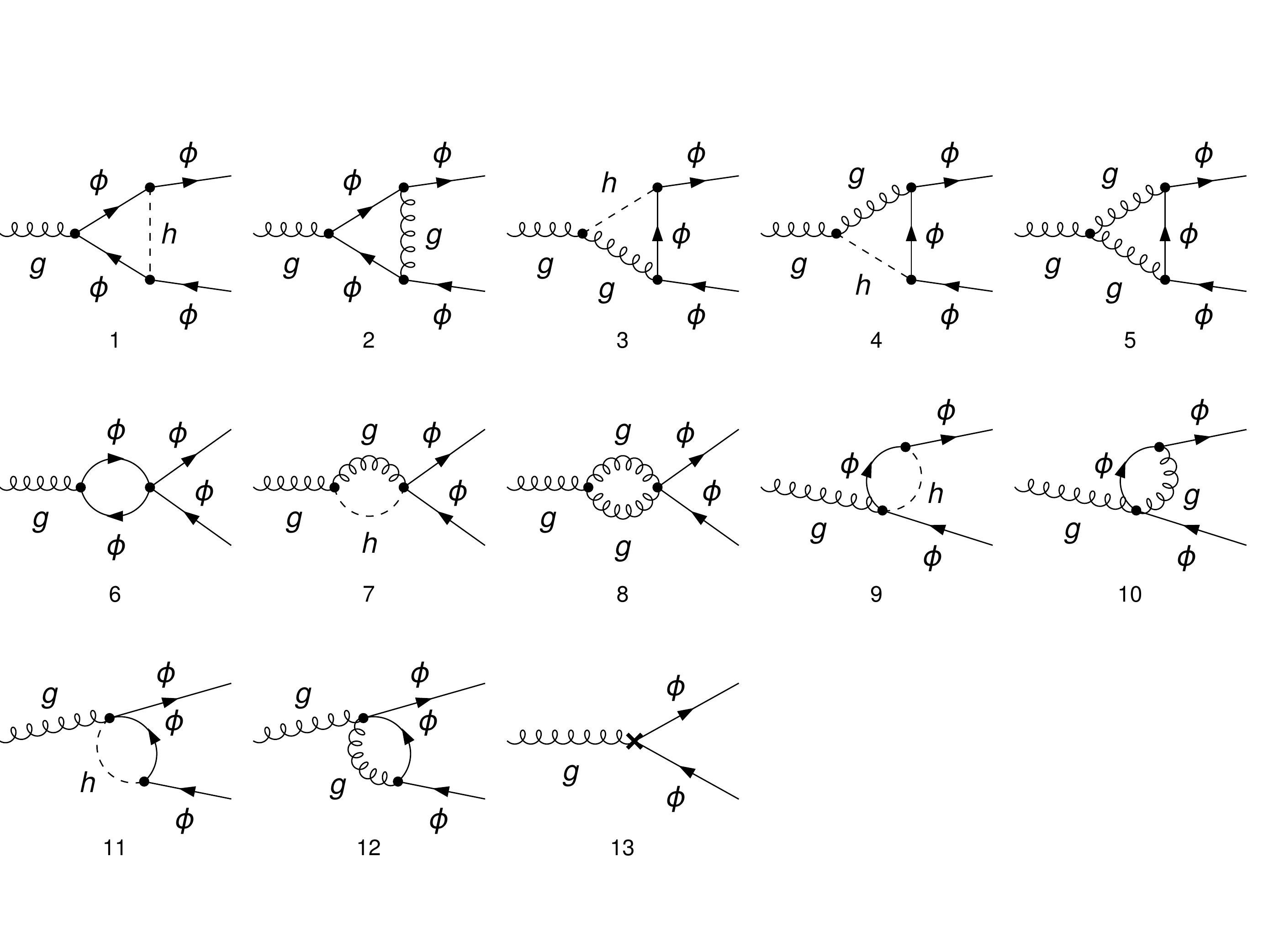}
	\caption{Feynman diagrams to the vertex interaction between quarks top and gluons up to one-loop order.}
	\label{fig05}
	\end{center}
\end{figure}

For the other 3-point function, the scalar-scalar-gluon vertex, the gravitational interaction will be present in some diagrams, as we can see in Fig.~\ref{fig05}, where the relevant contributions to this function up to one-loop order are shown. The resulting expression is
\begin{eqnarray}
 -i\Gamma^\mu_{abc}&=&gf_{abc}(p^\mu_2-p^\mu_3)\left(\frac{C_A \left(9-5 \xi _A\right) g^2+4 \kappa ^2 m^2 \left(\xi _h-2\right)}{64 \pi ^2 \epsilon }-Z_1^{(1)}\right)\nonumber\\
 &&+ O(p^3) + \text{finite},
\end{eqnarray}

\noindent from which, through MS, we find
\begin{equation}\label{eq_z1}
 Z_1^{(1)}=\frac{C_A \left(9-5 \xi _A\right) g^2+4 \kappa ^2 m^2 \left(\xi _h-2\right)}{64 \pi ^2 \epsilon }.
\end{equation}

\begin{figure}[t!]
	\begin{center}
	\includegraphics[angle=0 ,width=12cm]{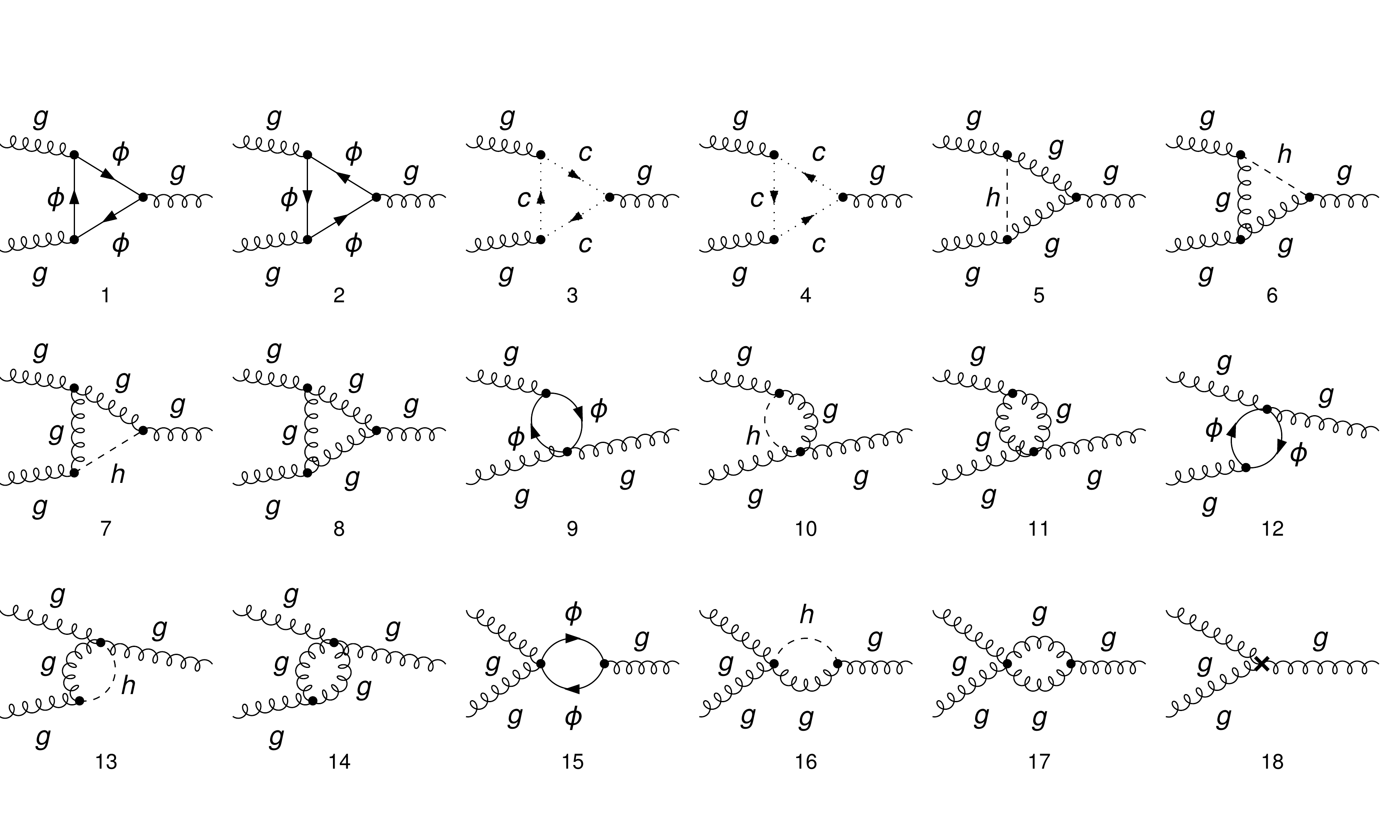}
	\caption{Feynman diagrams to the gluons vertex interaction at one-loop order.}
	\label{fig06}
	\end{center}
\end{figure}

The 3-point function describing the vertex with three gluons in shown in Fig.~\ref{fig06}. We have used the projection

\begin{equation}
 \Pi^{\mu\nu\alpha}_{abc}=\eta^{\mu\nu}\Pi^{\alpha}_{abc} \qquad \Rightarrow \qquad \Pi^\alpha_{abc} = \frac{1}{4}\eta_{\mu\nu}\Pi^{\mu\nu\alpha}_{abc}
\end{equation}
and used the fact that $p_3 = p_1 + p_2$, to get
\begin{eqnarray}
 -i\Pi^\alpha_{abc} &=& \frac{g^3 f_{abc}C_A\left(-9 \xi _{A}-4N_s+17\right)(p_1-p_2)^\alpha}{256\pi^2\epsilon}-\frac{3}{4} Z_{3g}^{(1)} g \left(p_1-p_2\right){}^{\alpha } f_{abc}\nonumber\\
 &&+O(p^2)+\mathrm{finite},
\end{eqnarray}
Through MS, we impose finiteness and find
\begin{equation}\label{eq_z13g}
 Z_{3g}^{(1)} = -\frac{g^2C_A \left(9\xi _A-17 -4N_s\right)}{192 \pi ^2 \epsilon }.
\end{equation}

Now, we consider the scattering of four gluons (Fig.~\ref{fig07} showed at the end of the paper for convenience). Since the interaction of four gluons has no derivatives, the $Z_{4g}$ counterterm will renormalize terms proportional to $p^0$ and therefore we can set external momentum equals to zero if we restrict ourselves to the computation of this counterterm. Also, for simplicity, we have used the scalar projection
\begin{equation}\label{p4g}
 \Gamma_{abcd} = \frac{1}{16}\eta_{\mu\nu}\eta_{\rho\sigma}\Gamma^{\mu\nu\rho\sigma}_{abcd},
\end{equation}
\noindent to obtain the expression for the gluon 4-point function
\begin{eqnarray}
 -i\Gamma_{abcd} &=& -\left(\frac{i C_A g^4 \left(N_s+3 \xi _{A}-2\right)}{32 \pi ^2 \epsilon }+\frac{3}{2} i Z_{4g}^{(1)} g^2\right)\Bigr( \text{tr} (t_{a}t_{b}t_{c}t_{d})-2 \text{tr} (t_{a}t_{c}t_{b}t_{d})-2 \text{tr}(t_{b}t_{c}t_{a}t_{d})\nonumber\\
 &&+ \text{tr}(t_{b}t_{a}t_{c}t_{d}) + \text{tr} (t_{c}t_{a}t_{b}t_{d})+\text{tr} (t_{c}t_{b}t_{a}t_{d})\Bigr),
\end{eqnarray}
\noindent Then, again imposing finiteness through MS, we have
\begin{equation}\label{eq_z14g}
 Z^{(1)}_{1_{4g}}=-\frac{C_A g^2 \left(N_s+3 \xi _{A}-2\right)}{48 \pi ^2 \epsilon }.
\end{equation}

The other 4-point function involves two scalars and two gluons (Fig.~\ref{fig08}, again showed at the end of the paper for convenience). For this vertex, we use the following projection
\begin{equation}
\Pi^{\mu\nu}_{abcd}=\eta^{\mu\nu}\Pi_{abcd} \qquad \Rightarrow \qquad \Pi_{abcd} = \frac{1}{4}\eta_{\mu\nu}\Pi^{\mu\nu}_{abcd}
\end{equation}
and then we have
\begin{eqnarray}
 \Pi_{abcd}&&=\left(\frac{i g^2-3 C_A \left(\xi _A-1\right) g^2-2(\xi_h-2) \kappa ^2 m^2}{16 \pi ^2 \epsilon }-2 i Z_{2g}^{(1)} g^2\right)\Bigr( 2\text{tr} (t_{a}t_{b}t_{c}t_{d})- \text{tr} (t_{a}t_{c}t_{b}t_{d})\nonumber\\
 &&- \text{tr}(t_{b}t_{a}t_{c}t_{d})- \text{tr}(t_{b}t_{c}t_{a}t_{d}) - \text{tr} (t_{c}t_{a}t_{b}t_{d})+2\text{tr} (t_{c}t_{b}t_{a}t_{d})\Bigr).
\end{eqnarray}
and the counterterm is found to be
\begin{equation}
 Z_{2g}^{(1)}=-\frac{3 C_A \left(\xi _A-1\right) g^2-2(\xi_h-2) \kappa ^2 m^2}{32 \pi ^2 \epsilon }.
\end{equation}

From Eqs.~\eqref{eq_z2f}, \eqref{eq_z3}, \eqref{eq_z2c}, \eqref{eq_z1c}, \eqref{eq_z1}, \eqref{eq_z13g}, \eqref{eq_z14g} we conclude that
\begin{equation}
 Z_1^{(1)}-Z_{2s}^{(1)} = Z_{3g}^{(1)} - Z_3^{(1)} = \frac{1}{2}\left(Z_{4g}^{(1)} - Z_3^{(1)}\right) = \frac{1}{2}\left(Z_{2g}^{(1)}-Z_{2s}^{(1)}\right) = Z_{1c}^{(1)} - Z_{2c}^{(1)} = -\frac{C_A g^2(3+\xi_A)}{64 \pi^2 \epsilon }
\end{equation}
so the Slavnov-Taylor identities \cite{Slavnov:1972fg,Taylor:1971ff} are indeed respected and thus gravitational interaction does not spoil the gauge symmetry. This result allows us to define a global color charge.

Moreover, we can show that the beta function is independent of $\kappa$ and $m$, as the expression the one-loop beta function of the color charge can be found through the relations between the renormalized coupling constants and the counterterms given by
\begin{subequations}\label{g0}
\begin{eqnarray}\label{eq_e_0}
g&=&\mu^{-2\epsilon}\frac{Z_{2s} Z_3^{1/2}}{Z_1}g_0;\\
g&=&\mu^{-2\epsilon}\frac{Z_{3}^{3/2}}{Z_{3g}}g_0;\label{g0Z3g}\\
g&=&\mu^{-2\epsilon}\frac{Z_{3}}{Z_{4g}^{1/2}}g_0;\label{g0Z4}\\
g&=&\mu^{-2\epsilon}\frac{Z_{2c}Z_3^{1/2}}{Z_{1c}}g_0;\\
g&=&\mu^{-2\epsilon}\frac{Z_{2}^{1/2}Z_{3}^{1/2}}{Z_{2g}^{1/2}}g_0.
\end{eqnarray}
\end{subequations}

Therefore, the beta function for the color charge is
\begin{eqnarray}
 \beta(g) &=& \lim_{\epsilon\rightarrow0}\mu\frac{dg}{d\mu}=\lim_{\epsilon\rightarrow0}\mu\frac{d}{d\mu}\left[g_{0}\left(1-Z_1^{(1)}+Z_{2s}^{(1)}+\frac{Z_3^{(1)}}{2}\right)\mu^{-2\epsilon}\right]\nonumber\\
 &=& -\frac{g^3}{(4\pi)^2}\left(\frac{11}{3}C_A-\frac{2}{6}N_s\right).
\end{eqnarray}

The observed outcome is gauge-independent, a characteristic that was previously established via a functional approach in Ref.\cite{Folkerts:2011jz}. This property has also been verified in the context of the Effective Field Theory of gravity when coupled with fermionic QCD in \cite{Souza:2022ovu}.

As we can see, it does not depend on the mass, so our choice to make all masses the same does not affect our result for the beta function at one-loop order. On the other hand, as discussed in \cite{Bevilaqua:2021uev}, at two-loop we would expect a $\sum_i \kappa^2 m_i^2$ term.

It is needed to stress here the importance of a regularization scheme that preserves the symmetries of the model. In fact, the authors in Ref.\cite{Folkerts:2011jz} showed that in the weak-gravity limit there is no gravitational contribution at one-loop order if the regularization scheme preserves the symmetries of the model, such as dimensional regularization. On the other hand, if the regularization scheme does not preserve all the symmetries, there will be a negative contribution to the beta function (as seen in \cite{Robinson:2005fj}).

\section{Two-loop Gluon self-energy}\label{sec4}

This section presents the computation of the two-loop gluon self-energy and its renormalization. TARCER \cite{Mertig:1998vk}, in combination with previously cited \textit{Mathematica} packages, is utilized for this computation. TARCER implements the Tarasov algorithm for the reduction of two-loop scalar propagator type integrals with external momentum and arbitrary masses \cite{Tarasov:1997kx}. The Feynman and harmonic gauges ($\xi_A = \xi_h = 1$) are used for simplicity, and the analysis is limited to the case in which there is only one scalar particle ($N_s = 1$).

The Feynman diagrams we need to compute are showed in Fig.~\ref{fig09}. Due to gauge invariance, our result can be expressed as
\begin{equation}
 \Pi^{(2)}_{\mu\nu} = \left(p^2g_{\mu\nu} - p_\mu p_\nu\right) \Pi^{(2)},
\end{equation}
\noindent where the function $\Pi^{(2)}$ is a scalar function that can be expressed in terms of a set of basic integrals. To present the results in a simplified manner, we will adopt a notation similar to the one used in the original TARCER paper \cite{Mertig:1998vk} for the basic integrals that will be utilized,
\begin{subequations}
 \begin{eqnarray}
  &&\textbf{A}_{\nu}(m) = \frac{1}{\pi^{D/2}}\int\frac{d^D k}{[k^2-m^2]^\nu}\\
  &&\textbf{B}_{\nu_1,\nu_2}(m_1,m_2) = \frac{1}{\pi^{D/2}}\int\frac{d^D k}{[k^2-m_1^2]^{\nu_1}[(k-p)^2-m_2^2]^{\nu_2}}\\
  &&\textbf{J}_{\nu_1,\nu_2,\nu_3}(m_1,m_2,m_3) = \frac{1}{\pi^D}\int\frac{d^Dk_1d^Dk_2}{[k_1^2-m_1^2]^{\nu_1}[k_5^2-m_2^2]^{\nu_2}[k_4^2-m_3^2]^{\nu_3}}\\
  &&\textbf{F}_{\nu_1,...,\nu_5}(m_1,...,m_5)=\frac{1}{\pi^D}\int\frac{d^Dk_1 d^Dk_2}{[k_1^2-m_1^2]^{\nu_1}[k_2^2-m_2^2]^{\nu_2}[k_3^2-m_3^2]^{\nu_3}[k_4^2-m_4^2]^{\nu_4}[k_5^2-m_5^2]^{\nu_5}},\nonumber\\
  &&~
 \end{eqnarray}
\end{subequations}
in which $p$ is the external momentum and we introduced $k_3 = k_1 - p$, $k_4 = k_2 - p$, and $k_5 = k_1-k_2$. 

Therefore, we can write
\begin{eqnarray}\label{two_loops_1}
 \Pi^{(2)} &=& c_1~\textbf{A}_{1}(m)~\textbf{B}_{1,1}(0,0) + c_2~\textbf{A}_{1}(m)~\textbf{B}_{1,1}(m,m) + c_3~\textbf{B}_{1,1}(0,0)~\textbf{B}_{1,1}(m,m) + c_4 \left(\textbf{A}_{1}(m)\right)^2\nonumber\\
 &&c_5 \left(\textbf{B}_{1,1}(0,0)\right)^2 + c_6 \left(\textbf{B}_{1,1}(m,m)\right)^2 + c_7~\textbf{J}_{1,1,1}(0,0,0) +c_8~\textbf{J}_{1,1,1}(m,m,0) + c_9~\textbf{J}_{2,1,1}(m,m,0)\nonumber\\
 &&c_{10}~\textbf{F}_{1,1,1,1,1}(0,m,0,m,m) + c_{11}~\textbf{F}_{1,1,1,1,1}(m,0,m,0,m).
\end{eqnarray}

All of the aforementioned integrals are established and can be found in Refs.\cite{Martin:2005qm,Martin:2003qz}, and the coefficients $c_i$ are presented in appendix \ref{apxA}. As we are only concerned with the renormalization of the gluon wave-function, we expand Eq.\eqref{two_loops_1} around $p=0$ and retain only terms proportional to $p^0$. Higher powers in the external momentum will be renormalized by higher-order terms. Thus, we obtain:
\begin{eqnarray}
\Pi^{(2)}&=&-\frac{i \lambda C_A~g^2 }{384 \pi ^4\epsilon}-\frac{i \kappa ^2 m^2 C_A~g^2}{256 \pi ^4\epsilon}+\frac{i C_A^2~g^4 \log \left(m^2\right)}{384 \pi ^4\epsilon}-\frac{i C_A^2~g^4 \log \left(-p^2\right)}{64 \pi ^4\epsilon}-\frac{i \lambda C_A~g^2}{384 \pi ^4\epsilon}+\frac{5 i \gamma  C_A^2~g^4}{384 \pi ^4\epsilon}\nonumber\\
&&+\frac{17 i C_A^2~g^4}{576 \pi ^4\epsilon}+\frac{5 i \log (4 \pi ) C_A^2~g^4}{384 \pi ^4\epsilon}+\frac{5 i C_A^2~g^4}{768 \pi ^4 \epsilon ^2} + O(p) + \mathrm{finite}.
\end{eqnarray}

Now, we should compute the 1-loop diagrams with counterterms insertion in Fig.~\ref{fig10}. By doing so, we obtain
\begin{eqnarray}
 \Pi^{(2)}_{\mu\nu CT} = (p^2g_{\mu\nu} - p_\mu p_\nu)\Pi^{(2)}_{CT},
\end{eqnarray}
where
\begin{eqnarray}
 \Pi^{(2)}_{CT} &=& -\frac{i C_A^2~g^4 \log \left(m^2\right)}{384 \pi ^4 \epsilon }+\frac{i C_A^2~g^4 \log \left(-p^2\right)}{64 \pi ^4 \epsilon }-\frac{5 i C_A^2~g^4}{384 \pi ^4 \epsilon ^2}+\frac{i \lambda  C_A~g^2}{192 \pi ^4 \epsilon }-\frac{5 i \gamma  C_A^2~g^4}{384 \pi ^4 \epsilon }\nonumber\\
 &&-\frac{59 i C_A^2~g^4}{2304 \pi ^4 \epsilon }-\frac{5 i \log (4 \pi ) C_A^2 ~g^4}{384 \pi ^4 \epsilon } + O(p) + \mathrm{finite}.
\end{eqnarray}

Therefore, we obtain that the two-loop gluon wave-function counterterm is given by
\begin{equation}
 Z_3^{(2)} = \frac{C_A^2 g^4}{256 \pi ^4 \epsilon }-\frac{5 C_A^2 g^4}{768 \pi ^4 \epsilon ^2}-\frac{\kappa ^2 m^2 C_A g^2}{256 \pi ^4 \epsilon }.
\end{equation}

\section{Concluding remarks}\label{summary}

In summary, we have evaluated the n-point functions for the Einstein-Scalar-QCD model and demonstrated that there are no gravitational corrections to the beta function of the color charge at one-loop order. Additionally, we have explicitly verified that the Slavnov-Taylor identities are preserved at this order of perturbation theory, indicating that the universality of the color charge is maintained. Lastly, we have computed the counterterm for the gluon wave-function at two-loop order.

It is important to contextualize our results and compare them with previous research. To this end, we will follow the discussion in \cite{Donoghue:2019clr} and highlight some distinctions between our findings and theirs. One such difference lies in the adoption of a distinct regularization scheme. In reference \cite{Tang:2008ah}, it is argued that there are three primary concerns that should be considered when working with quantum gravity: gauge invariance, gauge conditions introduced in the quantization process, and the ability of the method to regulate any type of divergence. It was further argued that although dimensional regularization (DR) satisfies the first two requirements, it cannot handle more than logarithmic divergences. Therefore, Tang and Wu employed the Loop Regularization method (LP) in their studies \cite{Tang:2008ah, Tang:2011gz} to regulate the divergences. This method is capable of dealing with the quadratic divergences that appear in the Feynman diagrams. The authors used LP to compute the beta functions of the Einstein-Yang-Mills theory and compared the results with those obtained using DR. They found that while using DR leads to no gravitational contribution at one-loop, the use of LP leads to a contribution that is proportional to $\mu^2$. 

It is a fundamental requirement that physical results should not depend on the choice of the regularization scheme. Anber pointed out in \cite{Anber:2010uj} that the quadratic divergences are not relevant when using the S-matrix, which is a physical quantity. Moreover, Toms demonstrated in \cite{Toms:2011zza} that it is possible to define the electrical charge in quantum gravity using the background field method in a physically meaningful way that is not influenced by the quadratic divergences. Therefore, such contributions should be regarded as unphysical and should not be included in the evaluation of the running coupling. 

An intriguing avenue for further investigation pertains to the existence of a non-Abelian scalar particle serving as a potential dark matter candidate, as well as the implications of quantum gravity for dark matter. In the study conducted in Ref.\cite{Calmet:2021iid}, the potential ramifications of quantum gravity on dark matter models were explored. It was demonstrated that quantum gravity would give rise to a fifth force-like interaction, setting a lower limit on the masses of bosonic dark matter candidates. The authors also argued that, due to the influence of quantum gravity, these potential candidates would decay. However, given the ongoing observation of dark matter in the present universe, the authors were able to calculate an upper bound on the mass of a scalar singlet dark matter particle. In our future work, we intend to investigate the mass range for a non-Abelian scalar dark matter candidate, as presented in our study. In such a scenario, the fifth force-like interaction would also be non-Abelian in nature. This particular scenario was discussed in \cite{Arkani-Hamed:2008hhe}.

In our future endeavors, we plan to investigate the dynamics of the renormalized coupling constant in non-Abelian gauge theories, considering the presence of fermions and scalars coupled to gravity at the two-loop level. This investigation will involve an expansion of our research to incorporate modified theories of gravity, such as quadratic gravity~\cite{Odintsov:1991nd,Salvio:2014soa,Donoghue:2018izj,Donoghue:2021cza}. Drawing on the qualitative analysis presented in \cite{Souza:2022ovu}, we expect that modified theories of gravity, characterized by unconventional properties such as repulsive gravity under specific regimes, could potentially impact the behavior of the beta function. These modified gravity theories introduce additional gravitational interactions and might influence the running of the coupling constant in non-Abelian gauge theories, leading to intriguing and novel phenomena.

\acknowledgments
The work of HS is partially supported by Coordena\c{c}\~ao de Aperfei\c{c}oamento de Pessoal de N\'ivel Superior (CAPES). 

\appendix
\section{Two-loop coefficients}\label{apxA}

 In this section we present the two-loop coefficients for the two-loop gluon self-energy from Eq.~\eqref{two_loops_1}.
 \begin{subequations}
  \begin{eqnarray}
   c_1 &=& -\frac{i \left(D^4-10 D^3+35 D^2-50 D+24\right) C_A g_s^2}{960 (D-4) (D-3) (D-1)^2 m^4} (-4 C_A g_s^2 (20 \left(2 D^2-3 D-11\right) m^2\nonumber\\
   &&+\left(2 D^2-11 D+12\right) p^2) -5 \left(D^2-8 D+12\right) \kappa ^2 m^2 \left((D-8) p^2-48 m^2\right));\\
   c_2 &=& -\frac{i C_A g_s^2}{16 (D-4) (D-3) (D-1)^2 m^2 p^2} (-64 (D-1)^2 \left(D^2-7 D+12\right) \lambda  m^2 \nonumber\\
   &&+8 (D-1) C_A g_s^2 \left(4 \left(D^3-8 D^2+19 D-16\right) m^2+(D-2) D p^2\right)+2 D^6 \kappa ^2 m^4-18 D^5 \kappa ^2 m^4\nonumber\\
   &&-D^5 \kappa ^2 m^2 p^2+22 D^4 \kappa ^2 m^4-64 D^4 \lambda  m^2+23 D^4 \kappa ^2 m^2 p^2+262 D^3 \kappa ^2 m^4+576 D^3 \lambda  m^2\nonumber\\
   &&-196 D^3 \kappa ^2 m^2 p^2-1124 D^2 \kappa ^2 m^4-1728 D^2 \lambda  m^2+696 D^2 \kappa ^2 m^2 p^2+8 D^2 \kappa ^2 p^4+1712 D \kappa ^2 m^4\nonumber\\
   &&+1984 D \lambda  m^2-1048 D \kappa ^2 m^2 p^2-24 D \kappa ^2 p^4-928 \kappa ^2 m^4-768 \lambda  m^2+544 \kappa ^2 m^2 p^2+16 \kappa ^2 p^4);\nonumber\\
   c_3 &=& -\frac{i \left(D^3-8 D^2+19 D-12\right) C_A g_s^2 \left(2 C_A g_s^2+\kappa ^2 \left(2 (D-2) m^2-(D-4) p^2\right)\right)}{2 (D-4) (D-3) (D-1)^2};\\
   c_4 &=& \frac{i \left(3 D^4-40 D^3+180 D^2-320 D+192\right) C_A g_s^2}{960 (D-6) (D-5) (D-4)^2 (D-3) (D-2) (D-1)^2 (3 D-4) m^4 p^4}  (-1920 (D-1)^2 (D^4-14 D^3\nonumber\\
   &&+71 D^2-154 D+120) \lambda  m^2 p^2 +4 \left(D^2-3 D+2\right) C_A g_s^2 (\left(2 D^3-19 D^2+54 D-45\right) (D-4)^2 p^4\nonumber\\
   &&+32 \left(4 D^5-48 D^4+113 D^3+616 D^2-3099 D+3470\right) m^4+4 (8 D^5-40 D^4-281 D^3+2224 D^2\nonumber\\
   &&-4899 D+3924) m^2 p^2)+5 (D-5) m^2 p^2 (\left(D^2-3 D+2\right) ((D^5-23 D^4+200 D^3-820 D^2+1584 D\nonumber\\
   &&-1056) \kappa ^2 p^2-384 \left(D^3-8 D^2+19 D-12\right) \lambda )+4 (5 D^7-113 D^6+1052 D^5-5122 D^4+13896 D^3\nonumber\\
   &&-20896 D^2+16032 D-4800) \kappa ^2 m^2));\\
   c_5 &=&  \frac{i C_A g_s^2}{128 (D-4) (D-1)^2}  (64 \left(D^3-5 D^2+2 D+2\right) C_A g_s^2+(-24 D^5+497 D^4-3680 D^3+12984 D^2\nonumber\\
   && -21560 D+11840); \kappa ^2 p^2)\\
   c_6 &=& \frac{i C_A g_s^2}{64 (D-4) (D-1)^2 p^2} (\kappa ^2 (16 \left(D^3-10 D^2+36 D-36\right) m^4-8 \left(D^3-10 D^2+48 D-48\right) m^2 p^2\nonumber\\
   &&+\left(D^3-10 D^2+64 D-64\right) p^4)-128 (D-1) C_A g_s^2 \left(2 m^2-p^2\right));\\
   c_7 &=& -\frac{i C_A g_s^2}{48 (D-6) (D-4)^2 (D-1) (3 D-4) p^2} (24 (9 D^6-189 D^5+1364 D^4-4756 D^3+9280 D^2\nonumber\\
   &&-10336 D+4992) C_A g_s^2+(6 D^8-35 D^7-2454 D^6+39327 D^5-240012 D^4+695044 D^3\nonumber\\
   &&-915664 D^2+366464 D+98304) \kappa ^2 p^2);\\
   c_8 &=& -\frac{i C_A g_s^2}{480 (D-4) (D-2) (D-1) m^2 p^4} (4 (D-2) C_A g_s^2 (32 \left(12 D^4-92 D^3-41 D^2+1577 D-2776\right) m^4\nonumber\\
   &&+4 \left(24 D^4-172 D^3+273 D^2+193 D-516\right) m^2 p^2+\left(6 D^4-67 D^3+271 D^2-468 D+288\right) p^4)\nonumber\\
   &&+5 \kappa ^2 m^2 p^2 (4 \left(6 D^6-213 D^5+2417 D^4-12716 D^3+34112 D^2-45272 D+23616\right) m^2\nonumber\\
   &&+\left(3 D^6-63 D^5+518 D^4-2092 D^3+4296 D^2-3968 D+1024\right) p^2));\\
   c_9 &=& \frac{i C_A g_s^2}{480 (D-4) (D-3) (D-2) (D-1) m^2 p^4} (4 (D-2) C_A g_s^2 (240 \left(7 D^2-57 D+100\right) m^4 p^2\nonumber\\
   &&-(D-4)^2 \left(2 D^2-9 D+9\right) p^6+128 \left(4 D^4-32 D^3-7 D^2+548 D-1041\right) m^6\nonumber\\
   &&-4 \left(6 D^4-39 D^3-22 D^2+517 D-876\right) m^2 p^4)+5 \kappa ^2 m^2 p^2 (16 (2 D^6-69 D^5+789 D^4-4236 D^3\nonumber\\
   &&+11684 D^2-16012 D+8664) m^4-4 (D^6-44 D^5+543 D^4-3040 D^3+8736 D^2-12616 D\nonumber\\
   &&+7296) m^2 p^2-\left(D^6-25 D^5+246 D^4-1220 D^3+3224 D^2-4416 D+2496\right) p^4));\\
   c_{10} &=& \frac{i \kappa ^2 m^2 C_A g_s^2 \left(\left(D^2-6 D+4\right) p^2-4 (D-2) m^2\right)}{2 (D-1)};\\
   c_{11} &=& -\frac{i C_A g_s^2\left(C_A g_s^2 \left(8 m^2-p^2\right)+(D-2) \kappa ^2 m^2 \left((D-4) p^2-8 m^2\right)\right)}{2 (D-1)}.
  \end{eqnarray}
 \end{subequations}

\newpage

\begin{figure}[h!]
	\begin{center}
	\includegraphics[angle=0 ,width=14.5cm]{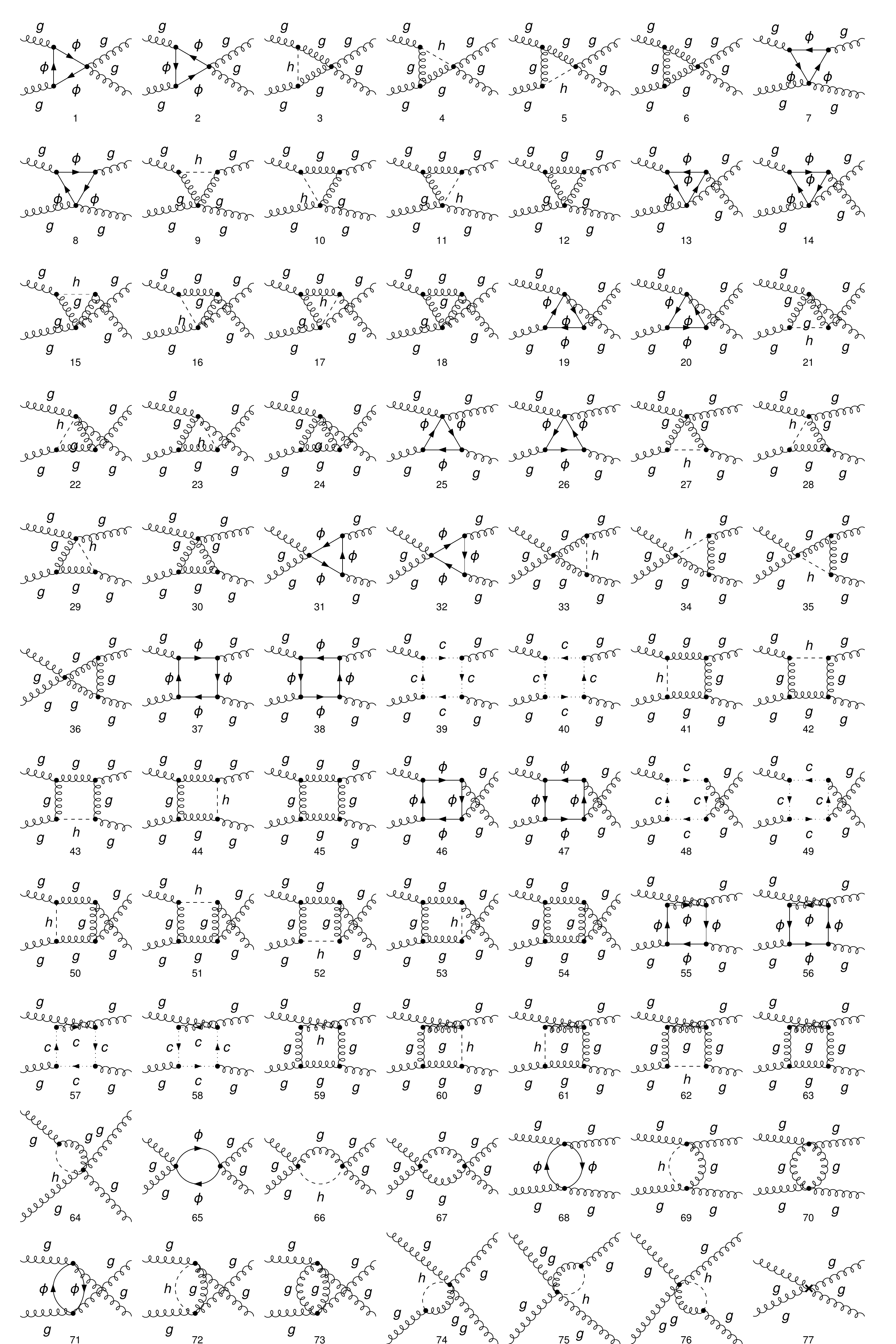}
	\caption{Feynman diagrams to the scattering between gluons up to one-loop order and one graviton exchange.}
	\label{fig07}
	\end{center}
\end{figure}

\begin{figure}[h!]
	\centering
	\includegraphics[angle=0 ,width=16cm]{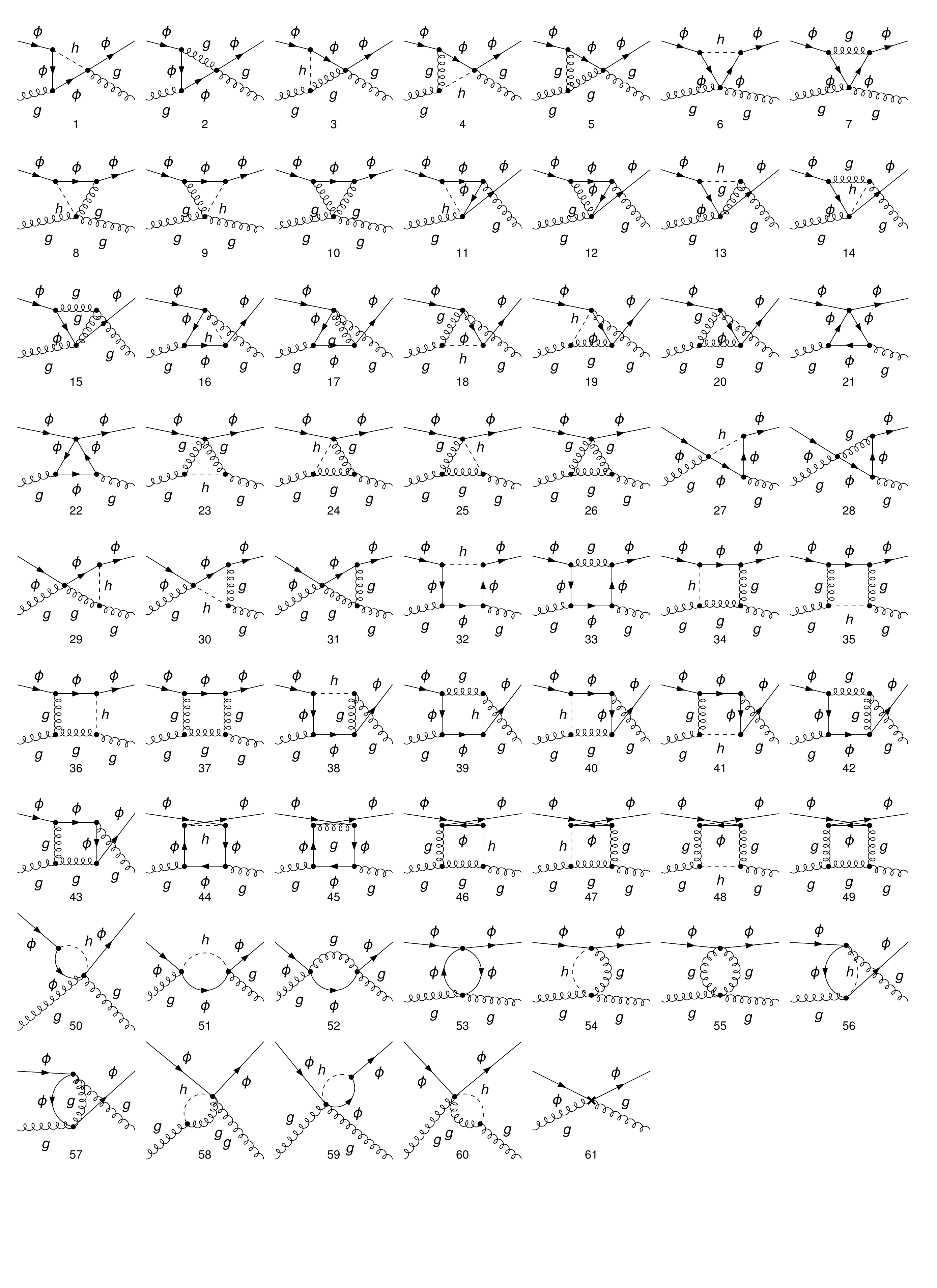}
	\caption{Feynman diagrams to the scattering between gluons and quarks up to one-loop order and one graviton exchange.}
	\label{fig08}
\end{figure}

\begin{figure}[h!]
	\centering
	\includegraphics[angle=0 ,height=24cm]{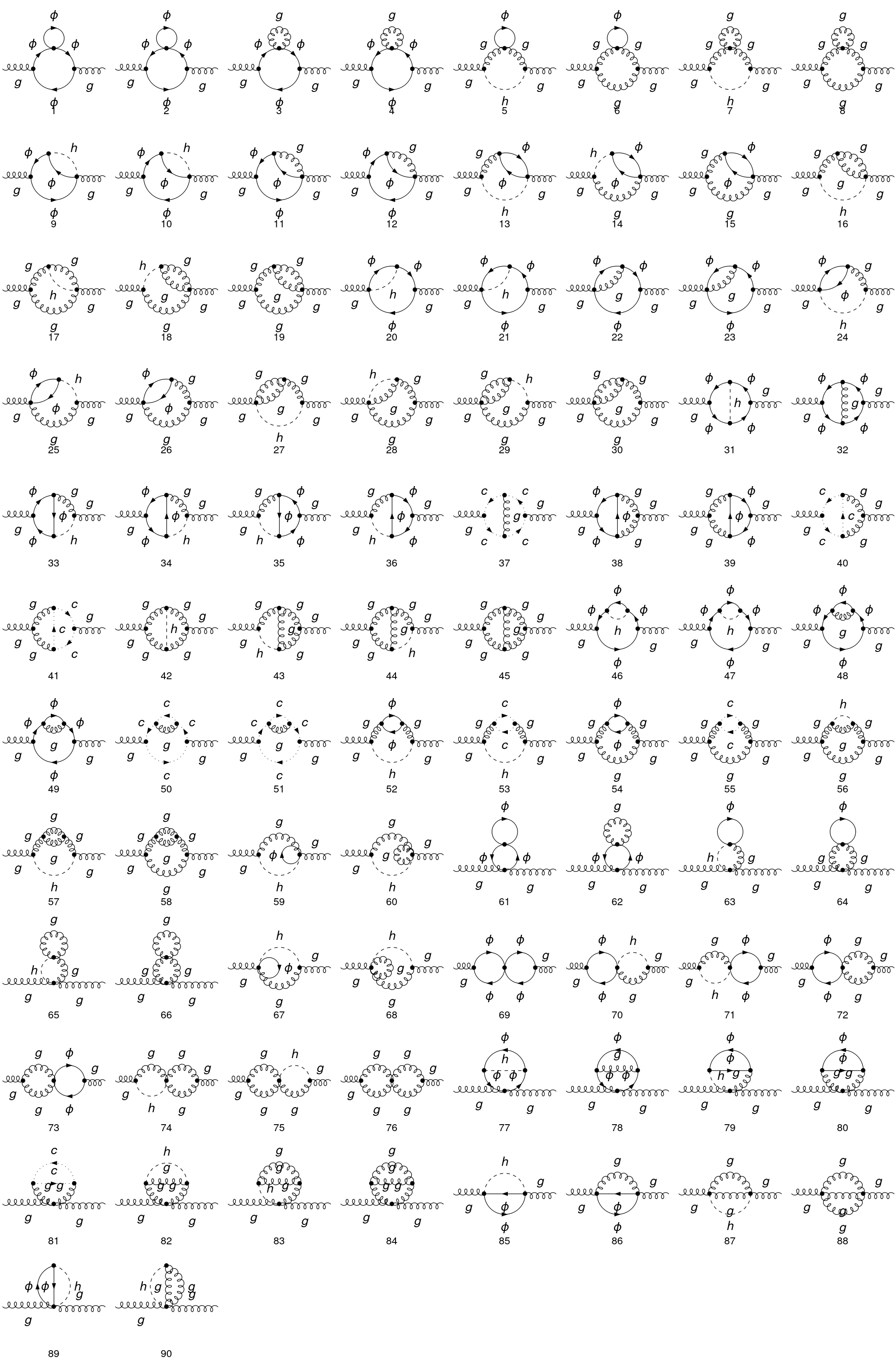}
	\caption{Feynman diagrams to the gluon self-energy involving only one graviton exchange at two-loop order.}
	\label{fig09}
\end{figure}

\begin{figure}[h!]
 \centering
 \includegraphics[angle=0,width=13cm]{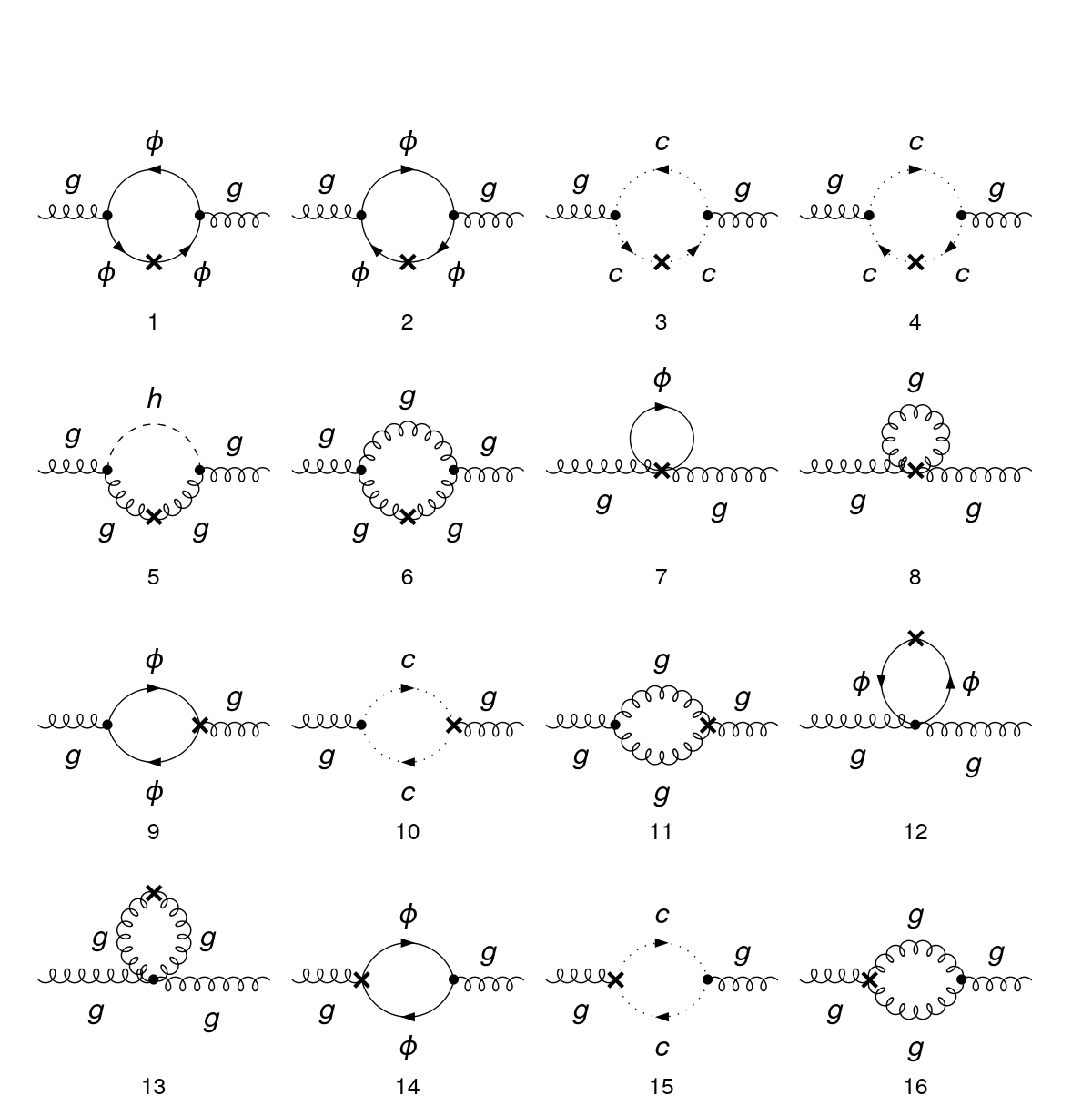}
 \caption{Gluon self-energy 1-loop diagrams with counterterms insertions.}
 \label{fig10}
\end{figure}

\end{document}